\documentclass[aps, prb, superscriptaddress, twocolumn, letterpaper, 10pt, amsfonts, amsmath, amssymb]{revtex4-2}

\usepackage{graphicx}
\usepackage{hyperref} 
\usepackage{xcolor}
\usepackage{soul}
\hypersetup{
    colorlinks,
    linkcolor={blue!80!black},
    citecolor={blue!80!black},
    urlcolor={blue!80!black}
}
\usepackage{physics}

\usepackage{placeins}

\usepackage{siunitx}
\usepackage{chemformula}
\begin{document}

\title{Topological superconductivity in semiconductor-superconductor-magnetic insulator heterostructures}

\author{A. Maiani}
\affiliation{Center for Quantum Devices, Niels Bohr Institute, University of Copenhagen, DK-2100 Copenhagen, Denmark}
\author{R. Seoane Souto}
\affiliation{Center for Quantum Devices, Niels Bohr Institute, University of Copenhagen, DK-2100 Copenhagen, Denmark}
\affiliation{Division of Solid State Physics and NanoLund, Lund University, S-22100 Lund, Sweden}
\author{M. Leijnse}
\affiliation{Center for Quantum Devices, Niels Bohr Institute, University of Copenhagen, DK-2100 Copenhagen, Denmark}
\affiliation{Division of Solid State Physics and NanoLund, Lund University, S-22100 Lund, Sweden}
\author{K. Flensberg}
\affiliation{Center for Quantum Devices, Niels Bohr Institute, University of Copenhagen, DK-2100 Copenhagen, Denmark}

\date{\today}

\begin{abstract}
Hybrid superconductor-semiconductor heterostructures are promising platforms for realizing topological superconductors and exploring Majorana bound states physics. Motivated by recent experimental progress, we theoretically study how magnetic insulators offer an alternative to the use of external magnetic fields for reaching the topological regime. We consider different setups, where: (1) the magnetic insulator induces an exchange field in the superconductor, which leads to a splitting in the semiconductor by proximity effect, and (2) the magnetic insulator acts as a spin-filter tunnel barrier between the superconductor and the semiconductor. We show that the spin splitting in the superconductor alone cannot induce a topological transition in the semiconductor. To overcome this limitation, we propose to use a spin-filter barrier that enhances the magnetic exchange and provides a mechanism for a topological phase transition. Moreover, the spin-dependent tunneling introduces a strong dependence on the band alignment, which can be crucial in quantum-confined systems. This mechanism opens up a route towards networks of topological wires with fewer constraints on device geometry compared to previous devices that require external magnetic fields.
\end{abstract}

\maketitle

\section{Introduction}
Topological superconductivity has been predicted to appear in one-dimensional spin-orbit coupled semiconductors proximitized by an \textit{s}-wave superconductor. In these systems, an external magnetic field causes the gap to close. In the presence of a strong spin-orbit interaction, the gap reopens, leading to a topological phase \cite{Oreg2010, Lutchyn_PRL2010}. The quantum phase transition brings the system into the topological regime when
\begin{equation}
    V_\mathrm{Z}^2 > \mu^2 + \Delta^2\,,
    \label{eq:nanowire_critical_lines}
\end{equation}
where $V_\mathrm{Z}$ is the Zeeman energy, $\mu$ the semiconductor chemical potential, and $\Delta$ the superconducting pairing amplitude. In the topological phase, the system behaves as a spinless \textit{p}-wave superconductor and Majorana bound states appear at the edges of the system. These states have been proposed to be used for topological quantum computation \cite{Kitaev_2001, Nayak2008}. Following this idea, different hybrid semiconductor-superconductor (Sm-Sc) platforms have been proposed to exhibit topological superconductivity, such as proximitized nanowires \cite{Lutchyn2018a}, selective-area-grown wires \cite{Vaitiekenas2018}, and two-dimensional electron gas (2DEG) systems \cite{Shabani2016,  Pientka_PRX2017, Hell2017, Fornieri_Nat2019}.

One limitation of these platforms is the requirement of external magnetic fields to induce the topological phase transition. External magnetic fields have several drawbacks as they are detrimental to superconductivity and the topological phase is very sensitive to the relative orientation between the magnetic and the spin-orbit fields. This problem becomes even more evident when considering more complicated geometries featuring several nanowires that cannot be aligned in the same direction, like the ones proposed to demonstrate Majorana non-Abelian statistics in real space \cite{Alicea_NatPhys2011, Aasen2016, Karzig2017}.

Magnetic materials are alternatives to the use of external magnetic fields. This idea was discussed by some early works in the field, where magnetic insulators (MIs) were used to induce a spin splitting by means of stray fields \cite{Fu2008, Sau2010}. In addition, clean interfaces with a MI also lead to exchange fields in the proximitized materials, which provide a more effective way to control the local spin splitting.

Developments in the fabrication technology have enabled the integration of thin layers of the MI \ch{EuS} in the hybrid Sm-Sc \ch{InAs}-\ch{Al} nanowires with excellent interface quality \cite{Liu2019, Blumel2019, Liu2020}. This material has also been tested in combination with \ch{Au} showing signatures consistent with the presence of Majorana bound states \cite{Manna2020}.

\begin{figure}[ht]
    \centering
    \includegraphics[width=1\columnwidth]{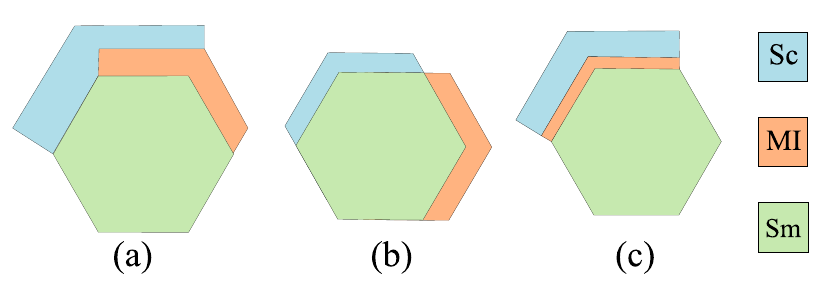}
    \caption{Sketch of the cross section of a Sm nanowire in proximity to a Sc and a MI.
    (a) Sm nanowire with partially overlapping Sc and MI, where a zero-bias peak has been recently reported \cite{Vaitiekenas2020a}. 
    (b) Sm nanowire with no overlap between the Sc and MI layers, which did not show signatures of topological superconductivity \cite{Vaitiekenas2020a}.
    (c) Sm nanowire with completely overlapping Sc and a thin MI layer. Measures of devices of the kind shown in (c) have not been reported yet. 
    }
    \label{fig:sketch_experiments}
\end{figure}

In experiments with ferromagnetic hybrid nanowires, spectroscopic measurements have shown the onset of a zero-bias conductance peak which has been interpreted as a signature of localized Majorana bound states at their ends \cite{Vaitiekenas2020a}. This signature has been detected only when the \ch{Al} and the \ch{EuS} layers overlap [Fig. \ref{fig:sketch_experiments}-(a)]. Samples with non-overlapping facets, like Fig. \ref{fig:sketch_experiments}-(b),  have shown no signatures of topological phases. This behavior is in contrast with expectations that the main effect of MIs is to induce an exchange field in the Sm \cite{Zhang2020}.

In these devices, however, the induced exchange field in the Sm is weak and short range \cite{Liu2020, Vaitiekenas2020a}. For this reason, the topological transition requires tuning the electrostatic gates, as pointed out by recent theory works \cite{Woods2020, Escribano2020, Liu2020a}. Another effect of the MI layer is to induce an exchange field in the Sc due to the proximity effect taking place at the Sc-MI interface \cite{Millis1988, Tokuyasu1988}. Many experimental works have verified a strong effect of MI on thin Sc layers in terms of an induced spin splitting of the Sc density of states \cite{Tedrow1986, Hao1990, Ren2007, Wolf2014, Strambini2017, Bergeret_RMP18}. Excluding a direct effect of the MI on the Sm, an alternative explanation can be that the spin-split Sc induces superconductivity and exchange field at the same time in the Sm. In this case, the relevant exchange coupling would be directly between the MI and the Sc \cite{Vaitiekenas2020a}. 

In this paper, we provide a theory for the combined superconductivity and magnetic proximity effect in Sc-MI-Sm and MI-Sc-Sm heterostructures, as illustrated in Fig. \ref{fig:sketch_theory}. We show that the combined magnetic and superconducting proximity effects of a spin-split Sc cannot induce a topological transition in the Sm by themselves. To overcome this limitation, we propose a new heterostructure layout where a thin film of MI between the Sm and the Sc leads to a spin-dependent tunnel barrier [Fig. \ref{fig:sketch_experiments}-(c)]. Our proposal exhibits a parameter region where topological superconductivity is present in the Sm for strong enough spin-dependent tunneling.

\section{Model}
\begin{figure}[ht]
    \centering
    \includegraphics[width=1\columnwidth]{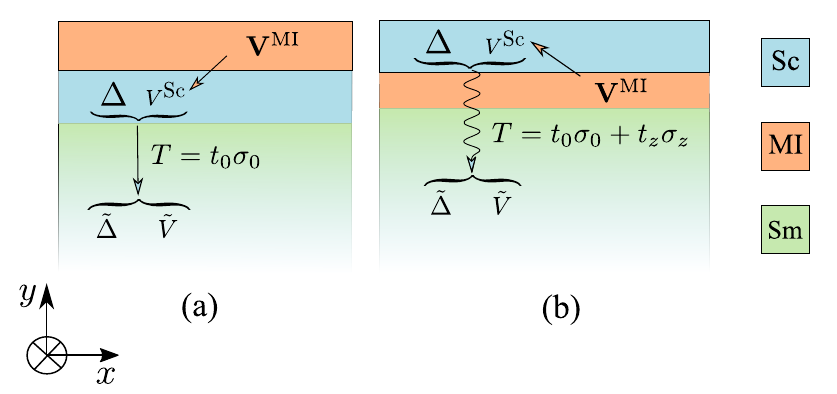}
    \caption{Sketch representing the proximity effects in the systems considered in this work. In the illustrations, $V^\mathrm{MI}$ is the native exchange splitting in the magnetic insulator and $\Delta$ and $V^\mathrm{Sc}$ are the gap parameter and the MI-induced exchange field in the Sc. Here, $\tilde{\Delta}$ and $\tilde{V}$ are the Sc-induced gap and exchange field in the Sm. The matrix $T$ is the hopping matrix of the tunneling Hamiltonian coupling the two materials.
    In panel (a), the magnetic insulator induces an exchange field in the Sc. The spin-split Sc is coupled to the Sm by a spin-symmetric tunneling Hamiltonian. In panel (b), a thin magnetic insulator layer is placed between the Sc and Sm. The MI induces an exchange field in the Sc, also providing a spin-dependent tunnel barrier between the Sc and the Sm. In both cases, we ignore the effect of the MI on the Sm as it only increases $\tilde{V}$ close to the interface to the MI.
}
    \label{fig:sketch_theory}
\end{figure}

A MI layer can induce several effects in the Sm-Sc device, depending on the heterostructure layout, as shown in Fig. \ref{fig:sketch_theory}.

One of the effects of the MI layer is the induction of an exchange field in both the Sm and the Sc. This is due to the microscopic scattering interaction between the electrons and the localized magnetic moments in the MI. We can describe the effect of the interface through a Heisenberg-like term $H_\mathrm{int} = -J \int \dd[3]{r} \vb{S}^\mathrm{MI} \cdot \vb{S}$, which describes the coupling between the spin density of the metals $\vb{S}$ and the localized spins in the insulator, $\vb{S}^\mathrm{MI}$. The coupling strength $J$ is related to the exchange integral between the localized orbitals and the free electron. This coupling strength is in principle different for the conduction band electrons of the MI and the electrons in the proximitized material \cite{Miao2015}. These considerations apply to ferromagnetic materials as well as antiferromagnetic insulators \cite{Kamra2018}. We consider the Sc layer width smaller than the superconducting coherence length, $\xi_0$, which is the characteristic decay length for the induced exchange field in a Sc-MI heterostructure \cite{Tokuyasu1988}. In the experiments \cite{Vaitiekenas2020a}, the SC and MI layers have a width of few nanometers while the typical size of a magnetic domain in the \ch{EuS} is $\sim10\xi_0$ for \ch{EuS}-\ch{Al} heterostructures \cite{Strambini2017}. Since the dimensions of the systems under consideration are smaller than $\xi_0$, we can disregard inhomogeneities and assume that the MI induces a homogeneous exchange field in the Sc, which couples to the spin of the electrons by the Zeeman term $H_\mathrm{Z} = \vb V^\mathrm{Sc} \cdot \vb*{\sigma}$. 

In this work we disregard the induced exchange field in the Sm as experimental evidence suggests that it is small in the tested devices \cite{Liu2020, Vaitiekenas2020a}. We note that recent theoretical works suggest that it can lead to topological superconductivity by careful control of the gate voltages \cite{Woods2020, Escribano2020, Liu2020a}. A direct MI-induced exchange field in the Sm would provide an enhancement of the Zeeman energy in the Sm, enlarging the topological region in our case. We also neglect the magnetic orbital effects induced by the stray field of the MI as they are usually weaker than ones induced by the exchange field \cite{Liu2019}.

If the MI considered is a magnetic semiconductor like \ch{EuS}, the conduction band is not accessible at low temperatures. The modest band gap can be used to fabricate spin filter tunnel barriers, using thin films, that allows spin-dependent tunneling through it when placed between two metallic regions \cite{Miao2015}. The proposed setup is illustrated in Fig. \ref{fig:sketch_theory}-(b).

In the following, we compare the two different situations represented in Fig. \ref{fig:sketch_theory}. For the case shown in Fig. \ref{fig:sketch_theory}-(a), the MI only causes a spin splitting in the Sc density of states. This situation is relevant for the geometry shown in Fig. \ref{fig:sketch_experiments}-(a) where a thick MI layer is placed in between the two materials such that the tunneling is strongly suppressed \cite{Vaitiekenas2020a}. In Fig. \ref{fig:sketch_theory}-(b) we show a different situation where a thin layer of MI is placed in between the Sc and Sm working as a spin-filter tunnel barrier, having a weak effect on the Sm density of states. This case applies to the device geometry shown in Fig. \ref{fig:sketch_experiments}-(c) but could also be relevant in the case of Fig. \ref{fig:sketch_experiments}-(a) if the MI layer is sufficiently thin and, therefore, all the interfaces contribute to the tunneling processes.

To simplify the treatment, we consider a translation-invariant system along the longitudinal direction, $z$, consisting of a single-channel Sm coupled to a Sc. The complete Hamiltonian of the system reads
\begin{equation}
    H = H_\mathrm{Sm} + H_\mathrm{Sc} + H_\mathrm{t}\,.
    \label{eq:hamiltonian_total}
\end{equation}

The Sm is described by
\begin{equation}
\label{eq:hamiltonian_nw}
\begin{split}
H_\mathrm{Sm} = \sum_{p_z} \qty(\frac{p_z^2}{2m_\mathrm{Sm}^*} - \mu_\mathrm{Sm}) c_{p_z}^\dag \sigma_0 c_{p_z} 
\\ + p_z c_{p_z }^\dag \alpha_x \sigma_y  c_{p_z}\,,
\end{split}
\end{equation}
where we use the spinors $c_{p_z} = (c_{p_z \uparrow}$, $c_{p_z \downarrow})$ and $c_{p_z \sigma}$ is the electron annihilation operator in the Sm, while $\alpha_x$ is the spin-orbit coupling strength.

The bare Hamiltonian for the Sc is given by
\begin{equation}
\label{eq:hamiltonian_sc}
\begin{split}
H_\mathrm{Sc} = \sum_{n p_z} & \xi_{n p_z} a_{n p_z}^\dag \sigma_0 a_{n p_z} + a^\dag_{n p_z} \vb{V}^\mathrm{Sc} \cdot \vb*{\sigma} a_{n p_z} +\\
                        & \qty( a^\dag_{n p_z} \vb*{\Delta}_{n p_z} i \sigma_y a^\dag_{n -p_z} + \mathrm{H.c.})\,,
\end{split}
\end{equation}
where $a_{n p_z}$ is the electron annihilation operator for the mode $n$ in the Sc and $\xi_{n p_z} = \frac{p_z^2}{2m_\mathrm{Sc}} + \varepsilon_n - \mu_\mathrm{Sc}$. We neglect the possible superconductive interband coupling and we assume singlet pairing in the parent superconductor gap matrix $\vb*{\Delta}_{n p_z} = \Delta_{0,n p_z} \sigma_0$. As anticipated, our Sc model features a homogeneous exchange field $\vb{V}^\mathrm{Sc} = V_z \vb{e}_z$ induced by the nearby MI, which we consider to be aligned to the wire. 
A crucial information is the distribution of the transverse modes in the Sc, $\varepsilon_n$, which strongly depends on the device geometry. The energy separation between transverse modes can range from a value larger than the superconducting gap for very thin Sc film to zero for a bulk superconductor, where the modes above the gap form a continuum.

The Sc and Sm regions are coupled through a spin-dependent, momentum conserving, tunneling Hamiltonian
\begin{equation}
\label{eq:hamitlonian_t}
H_\mathrm{t} = \sum_{n p_z} \qty(c^\dag_{p_z} T_{n p_z} a_{n p_z} + \mathrm{H.c.})\,,
\end{equation}
where the hopping matrix $T_{n p_z}$ describes the electron tunneling processes taking place at the interfaces between the two materials. We write $T_{n p_z}$ in the basis of Pauli matrices $ T_{n p_z} = t_{0, n p_z}\sigma_0 + \sum_i t_{i, n p_z} \sigma_i$. In the following, we set $t_{x, n p_z} = t_{y, n p_z} = 0$ as they are negligible for a uniformly polarized MI layer. Moreover, we consider only the case of real and positive $t_{0, n p_z}$ and $t_{z, n p_z}$. This is justified as we are assuming that spin-orbit coupling is absent in the Sc. An extended derivation with all the terms can be found in Appendix \ref{sec:appendix_complex_amplitudes}.

The bare Sc retarded Green's function in the basis of time-reversed pairs reads:
\begin{equation}
\begin{split}
    G^{R}_\mathrm{Sc} (\omega, n, p_z) = \Big[(\omega + i 0^+)\tau_0\sigma_0  - \xi_{n p_z}\tau_z \sigma_0  \\ 
    - \vb*{\Delta}_{n p_z}\tau_x\sigma_0 - V_z \tau_0 \sigma_z \Big]^{-1}\,,
\end{split}
\label{eq:green_function}
\end{equation}
where the $\tau_i$ are the Pauli matrices in the particle-hole space. In this work we ignore any back action of the Sm on the Sc as the electron density in the Sm is orders of magnitude smaller than the one in the Sc.

The retarded Green's function of the Sm reads $G^{R}_\mathrm{Sm}(\omega, n, p_z) = \qty[\omega + i 0^+ - H_\mathrm{Sm}(p_z) - \Sigma(\omega, p_z)]^{-1} $ where the self-energy $\Sigma(\omega, n, p_z) = \sum_n T_{n p_z} G^{R}_\mathrm{Sc}(\omega, n, p_z) T_{n p_z}^{\dag}$ describes the coupling to the Sc. This allows us to write a low-energy effective model for the Sm. Since we focus on the the quantum phase transition characterized by the gap closing, we work with the effective Hamiltonian $H_\mathrm{eff} = H_\mathrm{Sm} + \tilde{H}_0$ where the induced Hamiltonian is $\tilde{H}_0 = \Sigma(\omega=0)$. We can split the effective Hamiltonian into three different contributions:
\begin{equation}
\begin{split}
\tilde{H}_0 = \sum_{p_z} c_{p_z}^\dag \tilde{\mu}(p_z) \sigma_0 c_{p_z} + \\
+ c^\dag_{p_z} \tilde{V}_z(p_z) \sigma_z c_{p_z} + \\
+ \qty( c^\dag_{p_z} \tilde{\vb*{\Delta}}(p_z) i \sigma_y c^\dag_{-p_z} + \mathrm{H.c.})\,,
\end{split}
\label{eq:induced_hamiltonian}
\end{equation}
with $\tilde{\vb*{\Delta}}(p_z) = \tilde{\Delta}_0(p_z) \sigma_0 + \sum_i \tilde{\Delta}_i (p_z) \sigma_i$.
These three terms describe the shift in the chemical potential, the induced exchange field and the induced superconducting gap matrix in the Sm. The explicit forms of these contributions are

\begin{equation}
    \tilde{V}^{(1)}_z(p_z) = \sum_n \frac{V_z (t_{0, n p_z}^2 + t_{z,n p_z}^2 ) }{\xi_{n p_z}^2 + \Delta_{0, n p_z}^2 - V_z^2}\,,
    \label{eq:induced_V1}
\end{equation}

\begin{equation}
    \tilde{V}^{(2)}_z(p_z) = \sum_n \frac{- 2 \xi_{n p_z} t_{z, n p_z} t_{0, n p_z} }{\xi_{n p_z}^2 + \Delta_{0, n p_z}^2 - V_z^2}\,,
    \label{eq:induced_V2}
\end{equation}

\begin{equation}
    \tilde{\Delta}_0 (p_z) = \sum_n \frac{\Delta_{0, n p_z}(t_{0, n p_z}^2 - t_{z, n p_z}^2)}{\xi_{n p_z}^2 + \Delta_{0, n p_z}^2 - V_z^2}\,,
    \label{eq:induced_Delta}
\end{equation}
where we have divided the induced Zeeman term into two contributions $\tilde{V}_z = \tilde{V}_z^{(1)} + \tilde{V}_z^{(2)}$. The first one is proportional to the splitting in the parent Sc \eqref{eq:induced_V1}, while the second one is linked to the spin-asymmetric tunneling amplitude of the barrier $t_z$ \eqref{eq:induced_V2}. If we include the energy dependence, a triplet component $\tilde{\Delta}_z$ is also present, vanishing for $\omega=0$. Therefore, only the singlet pairing $\tilde{\Delta}_0$ is important to describe the topological transition. We assume that the parent Sc pairing potential is constant along $p_z$, $\Delta_{0, n p_z} = \Delta_{0, n}$. This is justified since the length of the wire is comparable to the Sc coherence length. 
For thin films, $\Delta_{0, n}$ is approximately constant for each band, even if different from the bulk value \cite{Wojcik2015}. For this reason, we further simplify the model by assuming the gap to be equal for each transverse mode of the Sc, i.e. $\Delta_{0, n} = \Delta_0$.

\section{Results}

\subsection{Semiconductor symmetrically coupled to a spin-split superconductor}

In this section, we show that the combined superconducting and exchange proximity effects induced by the coupling to a spin-split Sc cannot induce, alone, a topological transition in the Sm. We consider a system like the one sketched in Fig. \ref{fig:sketch_theory}-(a), where MI induces a spin splitting in the Sc. The Sc, in turn, induces superconductivity and an exchange field in the Sm. To check the presence of topological phases, we calculate the ratio between the induced exchange field, $\tilde{V}_z$, and the gap, $\tilde{\Delta}_0$. The topological phases appear when the condition in Eq. \eqref{eq:nanowire_critical_lines} is met, which leads to a closing of the superconducting gap at $p_z=0$. For this reason, we focus on this point in momentum space in the following. A necessary condition to satisfy the inequality is having a gap polarization ratio $\abs{V_\mathrm{Z}/\Delta} > 1$.

Taking the ratio of Eqs. \eqref{eq:induced_V1} and \eqref{eq:induced_Delta} for $t_z=0$, we see that
\begin{equation}
\abs{\frac{\tilde{V}_z}{\tilde{\Delta}_0}} = \abs{\frac{V_z}{\Delta_0}}\,,
\label{eq:no_go_theorem}
\end{equation}
so the induced gap polarization ratio in the Sm is the same as in the Sc.

The gap polarization ratio $\abs{V^\mathrm{Sc}/\Delta_0}$ has to be less than unity in the parent Sc, otherwise superconductivity is suppressed in the whole device. In the case of a large homogeneous Sc, this ratio is limited by the stricter Chandrasekhar-Clogston bound, which dictates that a finite superconducting gap can only be maintained if $\abs{V^\mathrm{Sc}/\Delta_0} < 1/\sqrt{2}$ \cite{Chandrasekhar_APL62, Clogston_PRL62}. In thin Sc films, quantum confinement causes an increase of the superconducting pairing amplitude leading to a superconductive phase that survives under stronger exchange fields. However, the Chandrasekhar-Clogston limit still holds in terms of gap polarization ratio \cite{Wojcik2015}. Even below this limit, the gap parameter in a clean Sc subjected to strong exchange fields ceases to be spatially homogeneous \cite{Fulde1964, Larkin1964}. Therefore, it is not possible to obtain topological phases by coupling a spin-split Sc to a Sm.

This result is independent of the mode distribution in the Sc within the constant $\Delta_0$ and $V_z$ approximation
\footnote{
If we allow the pairing potential to be different in each subband, the induced gap polarization ratio could take any value in the range of $\qty[\min (V_z/\Delta_{0, n}), \max (V_z/\Delta_{0, n})]$ depending on the coupling terms $T_n$. For a gaped superconductor all the subbands close to the Fermi level  fulfill $\Delta_{0, n}>V_z$. With this constrain, it is still not possible to get topological superconductivity in a semiconductor coupled to a spin-split superconductor through a spin-symmetric barrier.
}.
The same result is found for a continuous flat density of transverse modes in the Sc where $\Delta_0$ is taken finite and constant for a wide range of energies being zero otherwise, as shown in Appendix \ref{sec:appendix_BCS_integration}. In this case, the Zeeman term can be weakly enhanced, but this effect is totally negligible in realistic systems. This result can be generalized to the case where a multimode Sm is coupled to a multimode Sc as discussed in Appendix \ref{sec:appendix_multimode_sm}.

\subsection{Spin-dependent tunneling}
We consider now the case of a spin-asymmetric tunneling between the Sc and the Sm, taken as momentum independent for simplicity and described by $T = t_{0} \sigma_0 + t_z \sigma_z$. As Eqs. (\ref{eq:induced_V1})--(\ref{eq:induced_Delta}) show, the induced terms in the effective Hamiltonian are dependent on the distribution of the transverse modes in the Sc with respect to the chemical potential. In particular, they decay with the energy difference between the bottom of the Sc subband and the chemical potential. This means that the modes closer to the Sm Fermi energy give the dominant contribution to the induced superconducting pairing and exchange fields at $p_z=0$.

\begin{figure}[ht]
    \centering
    \includegraphics[width=1.0\columnwidth]{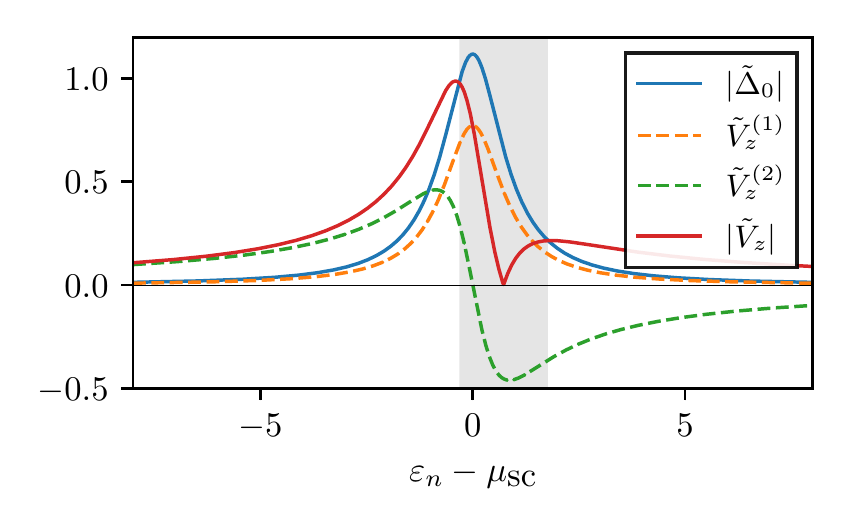}
    \caption{Induced gap $\tilde{\Delta}_0$ (blue line), exchange field contributions $\tilde{V}_z^{(1)}$ and $\tilde{V}_z^{(2)}$ (dashed lines), and total spin splitting $\tilde{V}_z=\tilde{V}_z^{(1)}+\tilde{V}_z^{(2)}$ (red line) at $p_z=0$ induced in the Sm. We show results for a single transverse mode of a spin-split Sc as a function of the chemical potential in the Sc, $\mu_\mathrm{Sc}-\varepsilon_0$.
    Parameters: $\Delta_0 = 1$, $t_0 = 1$, $t_z=0.4$, $V_z=0.5$. The area with gray background is topologically trivial while the one with white background satisfies the condition $\abs{\tilde{V}_z}>\tilde{\Delta}_0$}.
    \label{fig:single_mode}
\end{figure}

We first analyze the contribution of an isolated Sc mode to the effective Hamiltonian, as the ones from the different modes add up. The behavior of the induced term in the effective Hamiltonian is illustrated in Fig. \ref{fig:single_mode}.
The two terms of the induced exchange field $V^{(1)}_z$ and $V^{(2)}_z$ sum constructively for $\mu_\mathrm{Sc}>\varepsilon_n$. Both $\tilde{V}_z$ and $\tilde{\Delta}_0$ decay for $\abs{\xi_{n, 0}}\to\infty$, leading to the existence of an optimal regime where both the induced exchange field and superconducting pairing are maximal and $\abs{\tilde{V}_z/\tilde{\Delta}_0}>1$. This is the ideal region for searching for topological superconductivity.

In general, a spin-dependent component $t_i$ in the tunneling matrix enhances the collinear component in the exchange field $\tilde{V}_i$, suppressing the non collinear components in $\tilde{\vb{V}}$ and the superconducting gap in the Sm. From Eqs. (\ref{eq:induced_V1})--(\ref{eq:induced_Delta}) we see that $\tilde{\Delta}$ and $\tilde{V}_z^{(1)}$ share the same dependence on the Sc band structure (they decay as $\sim \xi^{-2}$) while having a different prefactor which depends on the tunneling matrix. Therefore, the spin splitting in the Sc leads to an enhancement of $\abs{\tilde{V}_z}$ and provides a first mechanism to induce topological superconductivity in the Sm.

In contrast to $\tilde{V}_z^{(1)}$, the second contribution $\tilde{V}_z^{(2)}$ to the induced exchange field in the semiconductor is totally independent of the spin polarization in the parent superconductor. This term depends solely on the spin-asymmetric component of the tunneling Hamiltonian. Moreover, $\tilde{V}_z^{(2)}$ has an energy dependence $\sim \xi^{-1}$, with a sign that depends on the relative position of the mode to the Sc Fermi energy. Since $\tilde{\Delta}_0$ and $\tilde{V}_z^{(1)}$ decay as $\sim \xi^{-2}$, $\tilde{V}_z^{(2)}$ dominates as the energy difference between the Sm and Sc modes increases. This contribution exhibits a sign change, which leads to a cancellation of  $\tilde{V}_z^{(2)}$ in the limit of small energy separation between modes. However, in thin Scs, the transverse modes can exhibit a large energy separation because of quantum confinement effects.

This result suggests that there are two different mechanisms that can drive the system to the topological phase. First, the spin-dependent tunneling enhances the induced gap-polarization ratio of spin-split superconductor. Alternatively, the spin-dependent tunneling combined with the quantum confinement in the Sc can lead to the appearance of the topological phase in regimes where, otherwise, the phase diagram would be globally trivial. 




\subsection{Combined proximity effect dominated by one transverse mode}
If the separation between transverse modes in the Sc is large enough, the contribution of the mode closest to the chemical potential dominates over the other ones. We can visualize this system like two coupled one-dimensional wires, where one features both spin splitting and superconductivity while the other features only spin-orbit coupling. In the rest of this section, we set the energy of the bottom of the single Sc band to zero, $\varepsilon_0 = 0$, as it only causes a shift of the energy scale. In this way $\mu_\mathrm{Sc} = -\xi_{0,p_z=0}$. 

\begin{figure}[ht]
    \centering
    \includegraphics[width=\columnwidth]{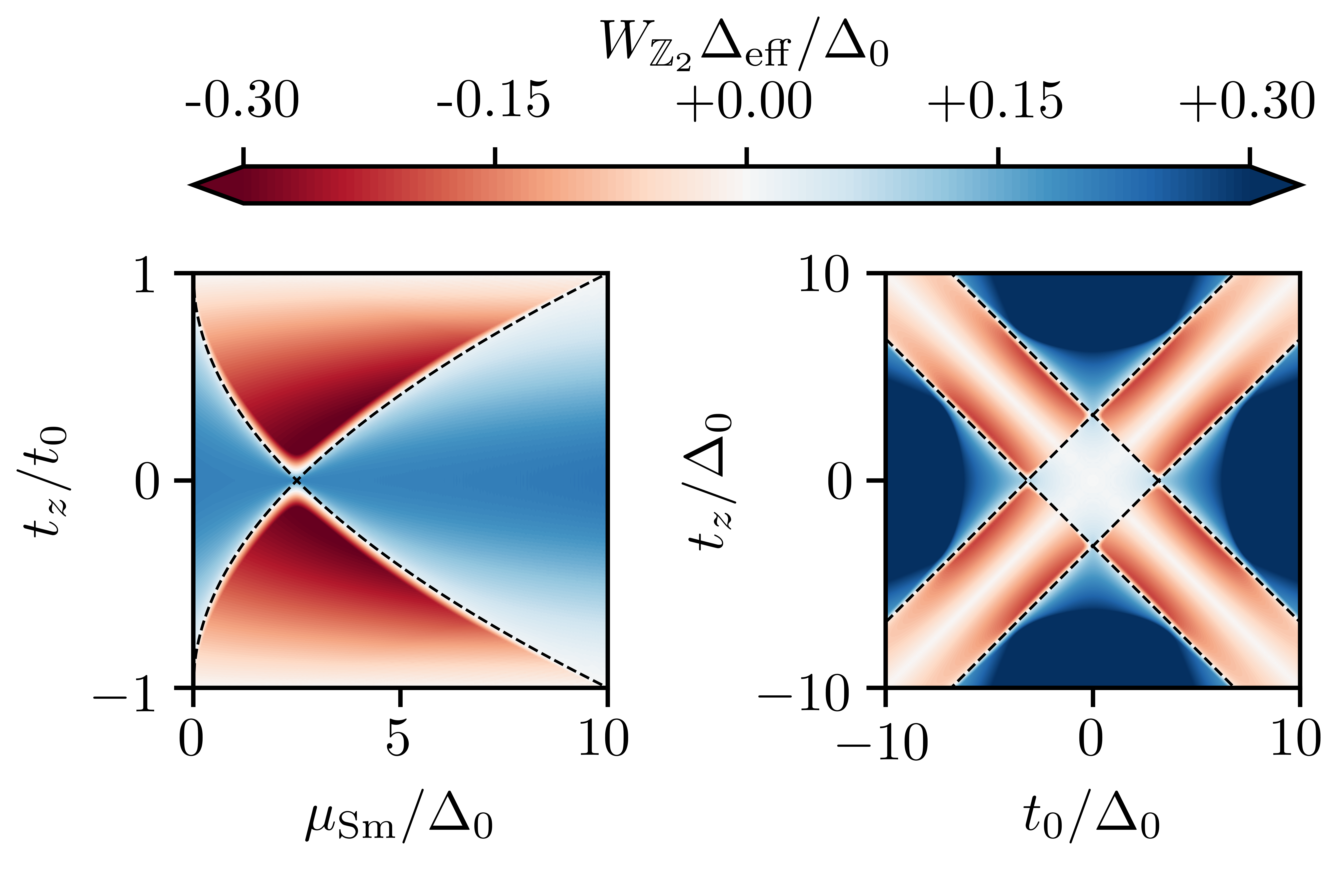}
    \caption{Phase diagram for a single mode Sm coupled to a single mode Sc. We represent the effective gap in the Sm $\abs{\Delta_\mathrm{eff}} = \min_{n, p_z} E_{n, p_z}$, where $E_{n, p_z}$ are the Hamiltonian eigenvalues and the sign is given by the topological invariant $W_{\mathbb{Z}_2}$ (Appendix \ref{sec:appendix_pfaffian_approach}). The black dashed lines demarcate the boundaries of the topological region (red color), described by Eqs. \eqref{eq:self_energy_critical_lines}. We have set  $V_{z} = 0$, which leads to $\tilde{V}_z^{(1)}=0$. The remaining parameters are $\mu_\mathrm{Sc}/\Delta_0 = 10$, $m^*_\mathrm{Sc} = 1$, $m^*_\mathrm{Sm} = 0.25$, $\alpha_x/\Delta_0 = 5$. For the left panel we take $t_0 = 5$ while for the right panel $\mu_\mathrm{Sm}/\Delta_0 = 1$.
    }
    \label{fig:single_mode_phase_diagrams}
\end{figure}

The topological transition can occur when the induced exchange field is larger than the induced gap. In this specific case, both the spin splitting and the superconducting pairing are inherited by the parent Sc, while the effective chemical potential is the sum of the bare nanowire electrochemical potential and the renormalization term induced by the proximity effect. To identify the critical lines, we impose the condition $\tilde{V}_z^2 = (\mu_\mathrm{Sm} + \tilde{\mu})^2 + \tilde{\Delta}_0^2$. To simplify the calculation, we take the limit $\Delta_0 \to 0$ as the chemical potential of the Sc is usually much bigger than the gap parameter. In this way, we find the analytical conditions

\begin{equation}
    \begin{split}
        t_z = +t_0 \pm \sqrt{+V_z \mu_\mathrm{Sm} + \mu_\mathrm{Sm} \mu_\mathrm{Sc}}\,,\\ 
        t_z = -t_0 \pm \sqrt{-V_z \mu_\mathrm{Sm} + \mu_\mathrm{Sm} \mu_\mathrm{Sc}}\,.
    \end{split}
    \label{eq:self_energy_critical_lines}
\end{equation}
This result allows us to identify the regions in the parameter space where the topological phases exist. The same result can be obtained by calculating analytically the Pfaffian invariant of this system, as shown in Appendix \ref{sec:appendix_pfaffian_approach}. An illustration of the phase diagram of this simple system is shown in Fig. \ref{fig:single_mode_phase_diagrams}, where topological superconductivity appears for a relatively wide range of parameters. Except for very low tunneling amplitudes, topological superconductivity is present in the region where $t_z/t_0 \sim 1$, which is the point corresponding to one spin tunneling component being completely suppressed (perfect spin filter). A strong polarization of the barrier, however, tends to suppress the induced gap, closing it for $t_z/t_0=1$. The topological region can be enlarged by controlling the chemical potential of the Sm and Sc. The size of the topological  region depends on the product $ \mu_\mathrm{Sm} \mu_\mathrm{Sc}$, as shown in Eqs. \eqref{eq:self_energy_critical_lines}.

The first conclusion is that the effect of the spin splitting of the Sc is negligible for realistic parameters, as it appears summed to the Sc Fermi energy which is many orders of magnitude bigger [Eqs. \eqref{eq:self_energy_critical_lines}]. Analyzing the behavior with respect to the tunneling parameters, it can be noticed that there exists a region corresponding to $t_0 \sim \pm  \sqrt{-V_z \mu_\mathrm{Sm} + \mu_\mathrm{Sm} \mu_\mathrm{Sc}}$, where a small polarization of the barrier can induce a topological phase with a relatively big gap. Since the chemical potential in the nanowire is tunable using electrostatic gates, it is, in principle, possible to set the operating point of the device near this optimal point. 

\subsection{Superconductor with several relevant transverse modes}
In a more realistic scenario, the Sc features many transverse modes. In this case, Eqs. (\ref{eq:induced_V1})--(\ref{eq:induced_Delta}) explain how the induced terms in the Sm effective Hamiltonian are determined by the sum of the contributions of each mode. 
The relative energy of the transverse modes in the Sc strongly affects the gap polarization in the Sm, becoming a critical factor for the appearance of the topological phase. For the systems of interest, the dimensions of the Sc section vary from a few to hundreds of nanometers. For small length scales, the effect of quantum confinement separates the Sc subbands such that it is not possible to treat the density of state like a continuum. 
Both $\tilde{\Delta}_0$ and $\tilde{V}_z^{(1)}$ have a Lorentzian shape with a full width at half maximum $\Gamma = \sqrt{\Delta_0^2 - V_z^2}$. We can use $\Gamma$ as a reference to distinguish between discrete modes, $\delta\varepsilon \gg \Gamma$, and a continuum distribution of modes, $\delta\varepsilon \ll \Gamma$, where $\delta\varepsilon$ is the average separation of modes. 
Indeed, if the average energy separation between the transverse modes is $\delta \varepsilon \gtrsim \Gamma$, the resonance peaks do not overlap entirely and the resonance peaks of $V_z^{(2)}$ do not completely cancel out. In this case, the net exchange field experienced in the Sm will be due to the sum of the contribution of each transverse mode.  

\begin{figure}[ht]
    \centering
    \includegraphics[width=1.0\columnwidth]{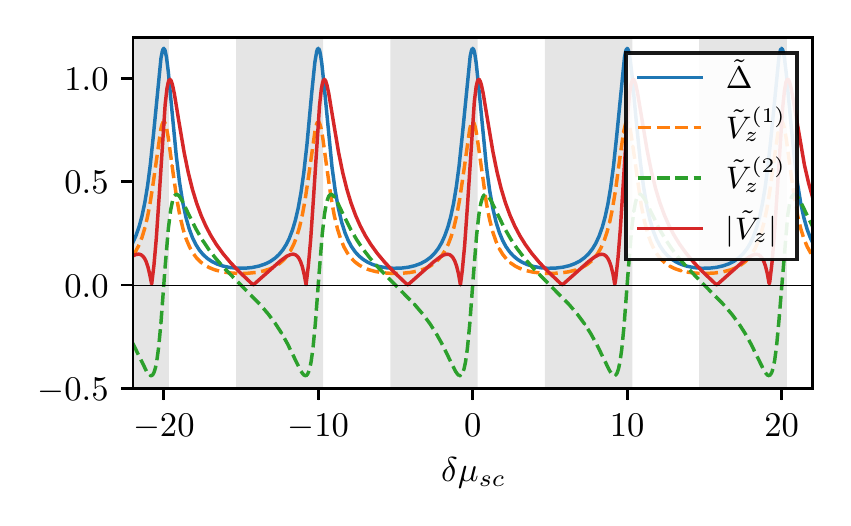}
    \caption{Induced gap $\tilde{\Delta}_0$ (blue line), exchange field contributions $\tilde{V}^{(1)}_z$ and $\tilde{V}^{(2)}_z$  (dashed lines) and total spin splitting $\tilde{V}_z=\tilde{V}_z^{(1)}+\tilde{V}_z^{(2)}$ (red line) at $p_z=0$ in the Sm in the case of a multimode Sc.
    Parameters: $\Delta_0 = 1$, $t_0 = 1$, $t_z=0.4$, $V_z=0.5$, $\delta\varepsilon=10$. The area with gray background is topologically trivial while the one with white background satisfies the condition $\abs{\tilde{V}_z}>\tilde{\Delta}_0$. The effect of the resonance peaks in $\tilde{V}_z^{(2)}$ cause an oscillation between topologically trivial and nontrivial phases while varying the chemical potential in the Sc.}
    \label{fig:discrete_modes}
\end{figure}

To provide a clearer picture, we consider the case where the band structure of the Sc can be described by perfectly equidistant transverse modes. In this case, $\delta\varepsilon$ is the relative separation between the Sc modes. This approximation recovers the continuum limit for a 2D system, where we expect a constant density of transverse modes $g_m = (\delta\varepsilon)^{-1}$. We also define $\delta\mu_\mathrm{Sc}$ as the relative position of the Fermi energy in the Sc from the middle of the band. We proceed by summing up the modes contributions following Eqs. (\ref{eq:induced_V1})--(\ref{eq:induced_Delta}) to calculate the value of the induced terms in the effective Hamiltonian (calculation details can be found in Appendix \ref{sec:appendix_multimode_sc}). The behavior of the induced terms for $p_z=0$ is illustrated in Fig. \ref{fig:discrete_modes}.

\begin{figure}[ht]
    \centering
    \includegraphics[width=1.0\columnwidth]{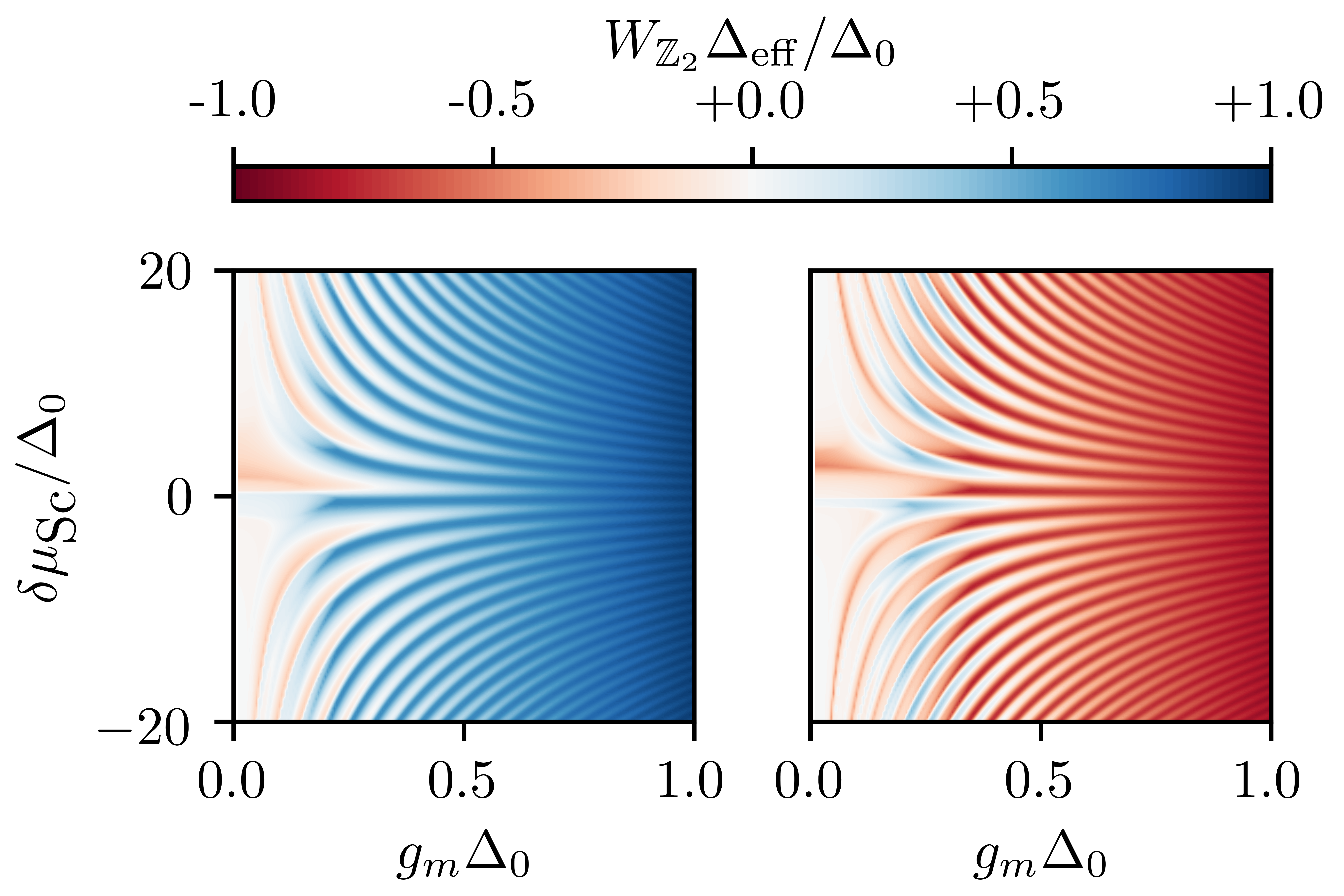}
    \caption{
    Phase diagram for a single mode Sm coupled to a multimode Sc, showing the effective gap in the Sm $\abs{\Delta_\mathrm{eff}} = \min_{n, p_z} E_{n, p_z}$, where $E_{n, p_z}$ are the Hamiltonian eigenvalues and sign given by the topological invariant $W_{\mathbb{Z}_2}$. We observe an oscillation between the topological (red) and trivial (blue) phases as we vary the chemical potential measured from the middle of the band $\delta\mu_\mathrm{Sc}$. As the density of transverse modes $g_m$ increases, the peaks get closer and overlap until they merge in the continuous limit $g_m \gg (\Delta_0)^{-1}$. In the continuum limit, the system is in a globally trivial or topological phase, depending on the condition in Eq. \eqref{eq:critical_point_continuum}. We assume that the chemical potential in the Sm is tuned such that $\mu_\mathrm{Sm} = - \tilde{\mu}$.
    Parameters used: $m^*_\mathrm{Sc} = 1$, $m^*_\mathrm{Sm} = 0.2$, $\alpha_x/\Delta_0 = 20$, $V_{z}/\Delta_0 = 0.5$, $t_0/\Delta_0=1$. For the left panel we use $t_z/\Delta_0 = 0.4$, while for the right panel we use $t_z/\Delta_0 = 0.7$.
    }
    \label{fig:multi_modes_phasediagram}
\end{figure}

By varying the chemical potential in the Sc, as the subbands cross the Fermi energy we see an alternation of trivial regions (gray background) and regions where topological superconductivity can appear by tuning $\mu_\mathrm{Sm}$ (white background). This behavior is more clearly illustrated by the topological invariant and the effective gap, shown in Fig. \ref{fig:multi_modes_phasediagram}. In this figure, the effective gap is given by the smallest eigenvalue of the effective Hamiltonian \eqref{eq:induced_hamiltonian}.

The observation of an alternation of topological and trivial phases while varying $\delta\mu_\mathrm{Sc}$ is consistent with Fig. \ref{fig:discrete_modes}. The limit $g_m \ll 1/\Delta_0$ corresponds to a large Sc where the effect of quantum confinement becomes completely negligible. This behavior can be realized if the average separation of the transverse modes is such that $\delta \varepsilon \ll \Gamma$. In this limit, the contributions from each mode overlap leading to a globally trivial or topological phase, depending on the polarization of the spin-filter barrier $t_z/t_0$. A quantitative criterion can be obtained by integrating Eqs. (\ref{eq:induced_V1})--(\ref{eq:induced_Delta}) over a constant density of transverse modes as shown in Appendix \ref{sec:appendix_BCS_integration}. In this limit, the term in Eq. \eqref{eq:induced_V2} vanishes and we get for the gap polarization
\begin{equation}
\abs{\frac{\tilde{V}_z}{\tilde{\Delta}_0}} = \abs{\frac{V_z}{\Delta_0}} \abs{\frac{t_0^2 + t_z^2}{t_0^2 -t_z^2}}\,.
\label{eq:ratio_with_tz}
\end{equation}

We note that the spin-dependent tunneling leads to an enhancement of the induced exchange field, while reducing the induced superconducting gap. This effect can be used to bring the induced gap polarization ratio above one, closing the gap at $p_z = 0$ and inducing a quantum phase transition to the topological phase. Therefore, the spin-asymmetric tunneling provides a way to overcome the limitation of Eq. \eqref{eq:no_go_theorem} and to induce a phase transition to the topological phase, Fig. \ref{fig:sketch_theory}-(b). The topological phase appears for a barrier polarization 
\begin{equation}
    \frac{t_z}{t_0} > \sqrt{\frac{1 - \abs{V_z/\Delta_0}}{1 + \abs{V_z/\Delta_0}}}\,.
    \label{eq:critical_point_continuum}
\end{equation}

\section{Discussion}
In this section, we discuss the applicability of the spin tunneling mechanism for generating topological superconductivity to the case of \ch{EuS}-\ch{Al}-\ch{InAs} platforms.  

\ch{EuS} is a magnetic semiconductor with an optical bandgap of $E_{g}=\SI{1.65}{\electronvolt}$, which can be effectively tuned by quantum confinement and, therefore, it depends on the film thickness \cite{Poulopoulos2012}. The magnetic properties of this material are attributed to the \ch{Eu} atoms which are characterized by strongly localized half-filled $4f$ shells. They behave as localized spins with a magnetic moment of around $7 \mu_B$. For this reason, the \ch{EuS} can be well described as a Heisenberg ferromagnet with a Curie temperature of $T_{m}=\SI{16.6}{\kelvin}$ \cite{Muller2002, Muller2004}. The magnetization induces a splitting of the conduction band of around $\Delta E_c=\SI{0.36}{\electronvolt}$ \cite{wachter1972optical}. These properties are very promising in the view of fabricating spin-filter tunnel barriers. 

As discussed in the previous section, the behavior of the system is strongly dependent on the band structure of the Sc, which is controlled by the dimension of the \ch{Al} region. If a wide \ch{Al} shell is used, the transverse modes of the nanowire will be densely distributed in the energy spectrum. In this case, $\tilde{V}_{z}^{(2)}$ vanishes and topological superconductivity can only be induced using the spin-asymmetric tunneling to enhance $\tilde{V}_z^{(1)}$ and suppress $\tilde{\Delta}_0$, Eq. \eqref{eq:ratio_with_tz}. 

To check if the ferromagnetic hybrid nanowire is in this regime, we can estimate the average mode separation by a simple particle in a box model $\delta\varepsilon \simeq \frac{\hbar^2}{2 m^*_\mathrm{Sc}}\frac{\pi^2}{ L_\mathrm{Sc}^2}$, where $L_\mathrm{Sc}$ refers to the largest dimension of the cross section of the Sc shell. Assuming that $\Gamma \sim \SI{100}{\micro\electronvolt}$, which is in line with the experimental measurements of Al-EuS heterostructures \cite{Strambini2017, Vaitiekenas2020a}, we estimate that, in order to observe well-separated modes, the maximum dimension of the shell should be in the order of \SI{60}{\nano\meter}. In experiments, the facet length is around \SI{60}{\nano\meter} \cite{Vaitiekenas2020a}. For this reason, we expect the contributions of the modes in the Sc to overlap significantly.

Previous measurements performed on \ch{EuS}-\ch{Al} heterostructures have shown a polarization of around 50\% of the gap \cite{Strambini2017}. In this case, applying \eqref{eq:critical_point_continuum}, we can estimate that a spin-dependent barrier with a 58\% polarization is enough to bring the Sm to the topological regime.

Estimating the optimal magnetic barrier length to achieve this polarization is a complicated task as strong spin-polarized band bending is expected at the interfaces. As the barrier gets thicker, we expect a stronger polarization, but at the same time, the coupling between the two systems gets strongly suppressed. Therefore, the optimal barrier length would be determined by the trade-off between a strongly polarizing thick barrier, which, however, suffers low transparency, and a thin transparent barrier with low polarization.

The introduction of MI in 2DEG is more challenging as the geometry and fabrication constraints make the introduction of Sc-Sm, MI-Sm, and MI-Sc interfaces at the same time particularly complicated. The use of MI as a spin-filter tunnel barrier simplifies the design of topological quantum devices by eliminating the need of the Sc-Sm interface. Using this new operating principle, the geometry sketched in Fig. \ref{fig:sketch_theory}-(b) can provide a viable alternative to achieve zero-field topological superconductivity in 2DEG devices.

Finally, we note that both nanowires Sc shell and Sc layers in 2DEG systems are around \SI{5}{\nano\meter} thick, so we can already expect a measurable effect of quantum confinement in this direction. For this reason, we expect that the resonance effect discussed previously can be measured by scaling down the width of the devices.

\section{Conclusions}
In this work, we have demonstrated that a spin-split superconductor cannot induce, alone, topological superconductivity in a spin-orbit coupled semiconductor by the combined superconducting and magnetic proximity effect. This is in contradiction with the hypothesis that an exchange field in the superconductor can induce a topological transition in the semiconductor.
We have shown how a spin-filter tunnel barrier, provided by a thin magnetic insulating layer between the semiconductor and the superconductor, can be used to overcome this limitation. The spin-dependent tunneling suppresses the induced superconducting pairing potential while enhancing the spin splitting, thus providing a way to bring the system to the topological phase. In the case of a distribution of discrete modes in the superconductor, the phase diagram exhibits an alternation of trivial and topological regions as a function of the chemical potential of the superconductor due to a change on the sign in the contributions to the exchange field. The total induced exchange field strongly depends on the relative energy difference between the closest levels to the Fermi energy in the superconductor. While this mechanism is unlikely to explain the results in Ref. \cite{Vaitiekenas2020a} because of the relatively thick \ch{EuS} layer used, the concept of spin-dependent coupling can be exploited in the next generation of topological superconducting devices without magnetic field. Spin filter tunnel barriers can be achieved by depositing a thin film of few nanometers of a ferromagnetic insulator at the interface between the superconductor and the semiconductor. The proposed mechanism is compatible with the currently used hybrid superconductor-semiconductor platforms, including nanowires and 2DEG systems.

\emph{Note added.} Recently, several independent papers on the same subject have been made publicly available as preprints. 

The result concerning the impossibility of topological phases by superconductor-mediated exchange field has been extended to generic systems in terms of a more general requirement about the minimal Zeeman field required for a topological phase transition \cite{Poyhonen2020}.

The ferromagnetic hybrid nanowire physics has been explored using a self-consistent treatment of a diffusive superconductor in Ref. \cite{Khindanov2020}. The results are in agreement with the ones presented in our work. Another independent work has found that perfectly clean systems with an extremely thin superconductive layer could be used to obtain topological phases in the MI-Sc-Sm stack \cite{Langbehn2020}.

\section{Acknowledgments}
This research was supported by the Danish National Research Foundation, the Danish Council for Independent Research | Natural Sciences, and the Microsoft Corporation. This project has received funding from the \textit{European Research Council (ERC) under the European Union's Horizon 2020 research and innovation programme} under Grant Agreement No. 856526. R.S.S. and M.L. acknowledge funding from QuantERA project “2D hybrid materials as a platform for topological quantum computing” and from NanoLund.
\bibliographystyle{apsrev4-2}

\bibliography{bibliography}

\clearpage
\appendix
\begin{widetext}
\section{Complete self-energy calculation}
\label{sec:appendix_self_energy_full}
In this appendix, we discuss more in detail the self-energy model employed to derive the effective Hamiltonian discussed in Eq. \eqref{eq:induced_hamiltonian}. We derive the model, discussing the frequency dependence, the presence of multiple modes in the Sm, and the effect of a general hopping matrix with arbitrary direction of the magnetization axis. Our model consists in a translation-invariant system composed by a Sm region coupled to a Sc in which an exchange field is induced by the nearby MI. This system is described by the Hamiltonian
\begin{equation}
    H = H_\mathrm{Sm} + H_\mathrm{Sc} + H_\mathrm{t}\,,
\end{equation}
where the three terms read
\begin{align}
\label{eq:h1}
    H_\mathrm{Sm} &= \sum_{m p_z} \zeta_{m}(p_z) c_{p_z}^\dag \sigma_0 c_{p_z} + \sum_{p_z} p_z c_{p_z }^\dag \alpha_x \sigma_y c_{p_z}\,,\\
\label{eq:h2}
    H_\mathrm{Sc} &= \sum_{n p_z} \xi_{n}(p_z) a_{n p_z}^\dag \sigma_0 a_{n p_z} + V_z a^\dag_{n p_z} \sigma_z a_{n p_z} + \qty[ a^\dag_{n p_z} (i \sigma_y \vb*{\Delta}_{n p_z}) a^\dag_{n -p_z} + \mathrm{H.c.}]\,,\\
\label{eq:h3}
        H_\mathrm{t} &= \sum_{m n p_z} c^\dag_{m p_z} T_{m n} (p_z) a_{n p_z} + \mathrm{H.c.}
\end{align}

We adopt the spinor basis $c_{p_z} = (c_{p_z \uparrow} c_{p_z \downarrow})$, where $c_{p_z \sigma}$ is the electron annihilation operator in the nanowire for spin $\sigma$. Both the Sm and the Sc have a multimode structure. In Eq. \eqref{eq:h1}, we label the Sm modes by the index $m$ and we write the kinetic energy as $\zeta_m(p_z) = \varepsilon_m + \frac{p_z^2}{2 m^*_\mathrm{Sm}} - \mu_\mathrm{Sm}$, where $m_\mathrm{Sm}^*$ and $\alpha_x$ are the electron effective mass and the spin-orbit coupling in the Sm. The Hamiltonian in Eq. \eqref{eq:h2} describes the Sc, where  $a_{n p_z}$ is the electron annihilation operator for mode $n$ with kinetic energy as $\xi_n(p_z) = \varepsilon_n + \frac{p_z^2}{2 m^*_\mathrm{Sc}} - \mu_\mathrm{Sc}$. Here, $\vb*{\Delta}_{n p_z}=\Delta_{0,n p_z} \sigma_0$ is the superconducting pairing for each band. We neglect superconducting interband coupling and the proximity-induced exchange field by MI in the Sc, $V_z$. The two regions are coupled by a spin-dependent tunneling Hamiltonian with hopping matrix $T_{m n} (p_z)$, which takes into account all the different electron transmission processes taking place at the interfaces between the different materials in the device. We can write $T_{m n p_z}$ in the basis of Pauli matrices:
\begin{equation}
    T_{m n}(p_z) = t_{0, mn}(p_z)\sigma_0 + \sum_{l=xyz}t_{l, mn} (p_z) \sigma_l\,,
\end{equation}
where the $t_{l, m n} = t_{l, m n}'(p_z) + i t_{l, m n}''(p_z)$ parameters are, in general, complex numbers.

In the following, we will use the basis of time-reversed pairs, where $\tau_i$ are Pauli matrices for the particle-hole space, being $\mathcal{H}$ the total Hamiltonian in this basis. In this basis, the bare Sc retarded Green's function reads 
\begin{equation}
\begin{split}
    G^{R}_\mathrm{Sc} (\omega, n, p_z) &=  \qty[(\omega + i 0^+)\tau_0\sigma_0 - \mathcal{H}_\mathrm{Sc}]^{-1} \\ &= \qty[(\omega + i 0^+)\tau_0\sigma_0 - V_z \tau_0 \sigma_z  - \xi_{n p_z}\tau_z \sigma_0 - \Delta_{n p_z }\tau_x\sigma_0 ]^{-1} = \\
    &= \frac{1}{(\omega + i 0^+ -  V_z \sigma)^2 - \xi_{n p_z}^2 - \Delta_{0, n p_z}^2} \qty[\qty(\omega + i 0^+ - V_z \sigma)\tau_0 + \xi_{np_z} \tau_z + \Delta_{n p_z}\tau_x]\,.   
\end{split}
\end{equation}

The full Green's function for the semiconductor subspace reads
\begin{equation}
G^R_\mathrm{Sm}(\omega, p_z) = \frac{1}{\omega + i 0^+ - \mathcal{H}_\mathrm{Sm} - \Sigma(\omega, p_z)}
\end{equation}
where $\mathcal{H}_\mathrm{Sm}$ is the bare Sm Hamiltonian matrix and we have defined the self-energy as
\begin{equation}
\qty[\Sigma(\omega, p_z)]_{m m'} = \sum_n \mathcal{H}_\mathrm{t, m n} (p_z) G^{R}_\mathrm{Sc}(\omega, n, p_z) \mathcal{H}_\mathrm{t,m' n}^{\dag}(p_z)\,,
\end{equation}
where the $\mathcal{H}_\mathrm{t}$ matrix is the tunneling Hamiltonian written in particle-hole space
\begin{equation}
    \mathcal{H}_\mathrm{t,mn} = (t_{0, mn}' (p_z) \tau_z + i t_{0, mn}''(p_z) \tau_0) \sigma_0 + \sum_l (t_{l, mn}' (p_z) \tau_0 + i t_{l, mn}'' (p_z) \tau_z)  \sigma_{l} \,.   
\end{equation}

In the next sections, we will explore in more detail different situations outlined in the main text adopting case-by-case assumptions and approximations.

\subsection{Frequency dependence}
\label{sec:appendix_frequency_dependence}
We now focus on the frequency dependence of the Sm Green's function. For this reason, we consider a single-band Sm, dropping the index $m$. We simplify the hopping matrix by assuming that the tunneling matrix has the following form $T_{n, p_z} = t_0 \sigma_0 + t_z \sigma_z$ with $t_0$ and $t_z$ real and positive. We proceed by splitting the self-energy like $\Sigma(\omega, p_z) = \tilde{E}(\omega, p_z) + \tilde{H}(\omega, p_z)$ where $\tilde{E} \propto \tau_0 \sigma_0$ while the frequency-dependent effective Hamiltonian $\tilde{H}(\omega, p_z)$ contains all the other $\tau_i \sigma_j$ terms. We study a low-energy model for the semiconductor valid in the limit $\omega\to0$. Since we have taken the tunneling amplitude energy independent, we can directly Taylor expand the Sc Green's function and consider the lowest orders in $\omega$:
\begin{equation}
\begin{split}
    G^{R}_\mathrm{Sc} (\omega, n, p_z) = &+ \frac{1}{V_z^2 - \xi_{n p_z}^2 - \Delta_{0, n p_z}^2} \qty[ - V_z \sigma_z \tau_0 + \xi_{n p_z} \sigma_0 \tau_z + \Delta_{0, n p_z}\sigma_0 \tau_x] + \\
    &+ \frac{\omega}{(V_z^2 - \xi_{n p_z}^2 - \Delta_{0, n p_z}^2)^2} \qty[ - (V_z^2 + \Delta_{0, n p_z}^2 + \xi_{n p_z}^2) \tau_0 \sigma_0 + 2 V_z \Delta_{0, n p_z} \sigma_z \tau_x + 2 V \xi_{n p_z} \sigma_z \tau_z] +  \\
    &+ O(\omega^2)\,.
\end{split}
\end{equation}

Introducing the expansion in the  self-energy term and keeping the zeroth and first order for each component, we see that the $\tau_0\sigma_0$ component reads
\begin{equation}
\tilde{E}(\omega, p_z) = - \omega \sum_n \frac{(t_0^2 + t_z^2) (\xi_{n p_z}^2 + \Delta_{0, n p_z}^2 + V_z^2) - 4 t_0 t_z V \xi_{n p_z}}{(\xi^2_{n p_z} + \Delta_{0, n p_z}^2 - V_z^2)^2} + O(\omega^2)\,.
\end{equation}

We further split the effective Hamiltonian according to $\tilde{H}(\omega, p_Z) = \tilde{H}_0 +  \tilde{H}_1(\omega, p_z) + O(\omega^2)$. The terms in $\tilde{H}_0$ have the same form as the ones in Eqs. (\ref{eq:induced_V1})--(\ref{eq:induced_Delta}) in the main text and do not depend on the energy. The first-order terms in $\omega$ are $\tilde{\mathcal{H}}_1(\omega, p_z) = \tilde{\Delta}_z(\omega, p_z) \tau_x \sigma_z + P(\omega, p_z) \tau_z \sigma_z$ where
\begin{equation}
\tilde{\Delta}_z(\omega, p_z) = - \sum_n \omega \frac{2 (t_0^2 - t_z^2) V_z \Delta_{0, n p_z} }{(\xi^2_{n p_z} + \Delta_{0, n p_z}^2 - V_z^2)^2}\,
\label{eq:InducedGapAppendix}
\end{equation}
and we denoted by $P(\omega, p_z)$ the remaining $\tau_z \sigma_z$ component. 

The semiconductor Green's function can be rewritten like
\begin{equation}
G_\mathrm{Sm}(p_z) = \frac{Z(p_z)}{\omega - Z(p_z) \qty(\mathcal{H}_\mathrm{Sm}(p_z) - \tilde{\mathcal{H}}_0(p_z) + \tilde{\mathcal{H}}_1(\omega, p_z))}\,,
\end{equation}
where $Z(p_z) =(1+ C(p_z))^{-1}$ with $\tilde{E} = - \omega C(p_z)$.

We can see that the first-order terms are proportional to $\omega/\sqrt{\xi_{n p_z}^2 + \Delta_{0, n p_z}^2 - V_z^2} < \frac{\omega}{\Gamma}$. Therefore, provided that we are interested in the $|\omega/\Delta_0|\ll1$ region, we can ignore these terms. This approximation breaks down in the strong-coupling regime, namely when the tunneling amplitudes overcomes $\Delta_0$. In this case, while the phase diagram can still be studied within the $\omega\to0$ limit, Eq. \eqref{eq:InducedGapAppendix} overestimates the size of the induced gap. For a more complete reference to the strong tunneling regime, we refer to \cite{Stanescu2017}.

\subsection{Multimode semiconductor}
\label{sec:appendix_multimode_sm}
Our objective now is to generalize the previous study to the case of a  multimode Sm in the case of spin-symmetric tunneling ($t_l=0$). Since we focus on the phase diagram, we consider the lowest order in the effective Hamiltonian $\tilde{\mathcal{H}}_0 = \Sigma(\omega=0, p_z)$. The three contributions read

\begin{equation}
    \tilde{\mu} (p_z) = \sum_n \xi_{n p_z} \frac{t_{0, m n p_z}t^*_{0, m' n p_z} }{\xi_{n p_z}^2 + \Delta_{0, n p_z}^2 - V_z^2}\,,
\end{equation}

\begin{equation}
    \tilde{V}^{(1)}_z(p_z) = \sum_n V_z \frac{t_{0, m n p_z} t^*_{0, m' n p_z}}{\xi_{n p_z}^2 + \Delta_{0, n p_z}^2 - V_z^2}\,,
    \label{eq:inducedVz1_multimode}
\end{equation}

\begin{equation}
\label{eq:inducedDelta_multimode}
\begin{split}
    \tilde{\Delta}_0 (p_z) = \sum_n [&+ (t_{0, m n p_z} t_{0, m' n p_z} - t_{0, m n p_z}' t_{0, m' n p_z}' ) \tau_x \\
    &- (t_{0, m n p_z}' t_{0, m' n p_z} + t_{0, m n p_z} t_{0, m' n p_z}' ) \tau_y ] \frac{\Delta_{n p_z}}{\xi_{n p_z}^2 + \Delta_{0, n p_z}^2 - V_z^2}\,.
\end{split}
\end{equation}

For $p_z=0$, $t_0$ real and $\Delta_{0, n p_z} = \Delta_0$, we can write the effective Hamiltonian matrix as 
\begin{equation}
    \mathcal{H}_\mathrm{eff} = \Xi \otimes \sigma_0 \tau_z + M \otimes ( V_z \sigma_z \tau_0 + \Delta_0 \sigma_0 \tau_x)
\end{equation}
where we have defined the $\Xi$ matrix as 
\begin{equation}
\Xi_{m m'} = \zeta_m \delta_{m m'} - \sum_n \xi_{n p_z} \frac{t_{0, m n p_z}t_{0, m' n p_z} }{\xi_{n p_z}^2 + \Delta_{0, n p_z}^2 - V_z^2} \,,  
\end{equation}
where $\zeta_m$ are the energies of the Sm modes. The $M$ matrix is 
\begin{equation}
M_{m m'} = \sum_n \frac{t_{0, m n p_z} t_{0, m' n p_z}}{\xi_{n p_z}^2 + \Delta_{0, n p_z}^2 - V_z^2}\,.
\end{equation}

The $M$ matrix is Hermitian so it is diagonalizable as $M=U\Lambda U^\dag$. We can apply the unitary transformation $U \otimes \tau_0 \sigma_0$ to the Hamiltonian matrix and get to
\begin{equation}
\begin{split}
\mathcal{H}_\mathrm{eff} &= U^\dag \Xi U \otimes \sigma_0 \tau_z + \Lambda \otimes (V_z \sigma_z \tau_0 + \Delta_0 \sigma_0 \tau_x) \\
    &= \hat{T} \otimes \sigma_0 \tau_z + \sum_m \qty[\hat{\zeta}_m \sigma_0\tau_z + \lambda_m (V_z \sigma_z \tau_0 + \Delta_0 \sigma_0 \tau_x) ] \delta_{m m'}\,,
\end{split}
\end{equation}
where the last term is the diagonal element $\hat{\zeta}_m$. The off-diagonal hopping terms in the matrix $\hat{T}$ come from the transformed $U^\dag \Xi U \otimes \sigma_0\tau_z$. This Hamiltonian is a combination of independent modes $\sum_m [\hat{\zeta}_m \tau_z\sigma_0 + \lambda_m (V_z \sigma_z \tau_0 + \Delta_0 \sigma_0 \tau_x) ]$ interacting through the spin-independent $\hat{T}$ terms which acts as a tunneling Hamiltonian for these channels. The gap polarization ratio is given by $\abs{V_z / \Delta_0}$ for these channels, which is the same as in the parent Sc. For this reason, if the parent Sc is in a trivial state, all the modes in the Sm will be in the trivial regime as well. The eigenmodes of the full effective Hamiltonian will be given by super-positions of these states. However, since we are mixing trivial states we can only obtain trivial states as a result. 

\subsection{Case of general tunnel matrix with complex element}
\label{sec:appendix_complex_amplitudes}
We now move to the case of a general tunneling matrix $T_{n p_z}$ coupling a multimode Sc to a monomodal Sm (thus we drop the index $m$). In this case we consider only the zeroth-order term in $\omega$ which is the effective Hamiltonian $\tilde{H}_0$. We can split it into three different contributions:
\begin{equation}
\tilde{H}_0 = \sum_{p_z} \tilde{\mu}(p_z) c_{p_z}^\dag\sigma_0 c_{p_z} + c^\dag_{p_z} \tilde{\vb{V}}(p_z) \cdot \vb*{\sigma} c_{p_z} + \qty( c^\dag_{p_z} \tilde{\vb*{\Delta}}(p_z) c^\dag_{-p_z} + \mathrm{H.c.})\,,
\end{equation}
with $\tilde{\vb*{\Delta}}(p_z) = \tilde{\Delta}_0(p_z) \sigma_0 + \tilde{\Delta}_i(p_z) \sigma_i$ and $\tilde{\vb{V}} = \tilde{\vb{V}}^{(1)} + \tilde{\vb{V}}^{(2)}$. The three terms are given by
\begin{equation}
    \tilde{\mu} (p_z) = \sum_n \frac{+\xi_{n p_z} (\abs{t_{0, n p_z}}^2 + \abs{t_{x, n p_z}}^2 + \abs{t_{y, n p_z}}^2 + \abs{t_{z, n p_z}}^2) - V_z \qty[(t_{0, n p_z} t_{z, n p_z}^* + \mathrm{H.c.}) + (t_{x, n p_z} t_{y, n p_z}^* - \mathrm{H.c.})]}{\xi_{n p_z}^2 + \Delta_{0, n p_z}^2 - V_z^2}\,,
\end{equation}

\begin{equation}
    \tilde{V}^{(1)}_z(p_z) = \sum_n V_z \frac{ \abs{t_{0, n p_z}}^2 + \abs{t_{z,n p_z}}^2 - \abs{t_{x,n p_z}}^2 - \abs{t_{y,n p_z}}^2}{\xi_{n p_z}^2 + \Delta_{0, n p_z}^2 - V_z^2}\,,
    \label{inducedVz1}
\end{equation}

\begin{equation}
    \tilde{V}^{(2)}_i(p_z) = \sum_n  - \xi_{n p_z} \frac{ \qty[(t_{0, n p_z} t_{i, n p_z}^* + \mathrm{H.c.}) + (t_{k, n p_z} t_{j, n p_z}^* - \mathrm{H.c.})]}{\xi_{n p_z}^2 + \Delta_{0, n p_z}^2 - V_z^2} \epsilon_{ijk}\,,
    \label{inducedVz2}
\end{equation}

\begin{equation}
    \tilde{\Delta}_0 (p_z) = \sum_n \frac{\Delta_{n p_z}(t_{0, n p_z}^2 - t_{x, n p_z}^2 - t_{y, n p_z}^2 - t_{z, n p_z}^2)}{\xi_{n p_z}^2 + \Delta_{0, n p_z}^2 - V_z^2}\,.
    \label{inducedDelta}
\end{equation}
where $\epsilon_{ijk}$ is the Levi-Civita symbol.
We can see that the presence of transverse components in the tunneling matrices causes a reduction of both the induced exchange field and superconducting pairing potential. However, if these amplitudes are complex, a non trivial resonance contribution can appear in the $\tilde{V}_{i}^{(2)}$ term enhancing the induced exchange field. Moreover, when complex amplitudes are present the induced gap can show interference effects due to the coupling between the various Sc bands induced by the Sm. 

\section{BCS integration induced gap and exchange field}
\label{sec:appendix_BCS_integration}
In this section we integrate the induced gap and exchange field due to the coupling to the superconductor. We take the usual BCS approximation which considers an attractive interaction between electrons with absolute energies below $\omega_D$. Therefore, the parent gap becomes finite and equal to $\Delta_0$ for energies in the range $[-\omega_D,\omega_D]$, being zero otherwise. Using this approximation, the induced gap is given by
\begin{equation}
    \tilde{\Delta}_0 = 2\Delta_0\,g_m(t_{0}^2-t_{z}^2)\frac{\mbox{atan}\left(\omega_D/\sqrt{\Delta^{2}_0-V^{2}_z}\right)}{\sqrt{\Delta^{2}_{0}-V_{z}^2}}\,,
\end{equation}
where $g_m$ is the density of transverse modes, considered to be energy independent. The induced exchange field is given by
\begin{equation}
    \tilde{V}_z = V_z\,g_m(\abs{t_{0}}^2+\abs{t_{z}}^2)\frac{\pi}{\sqrt{\Delta^{2}_0-V^{2}_z}}
\end{equation}
where the superconductor bandwidth has been taken infinite for simplicity.

For $t_z=0$, $\tilde{V}_z/V_z\geq\tilde{\Delta}_0/\Delta_0$, where the equal case corresponds to the limit $\omega_D\to\infty$. Taking realistic numbers for bulk aluminum, where $\omega_D/\Delta_0\approx200$, we find $(\tilde{V}_z/V_z-\tilde{\Delta}_0/\Delta_0)/(\tilde{\Delta}_0/\Delta_0)\sim10^{-2}$. If we consider that the Sc proximitized by the MI fulfills the Chandrasekhar-Clogston limit, $V_z<\Delta_0/\sqrt{2}$ \cite{Chandrasekhar_APL62,Clogston_PRL62}, the spin splitting cannot cause a topological transition in the Sm, unless $\omega_D/V_z\sim1$. We do not expect that an energy-dependent tunneling amplitude significantly changes this conclusion.

\section{Analytic derivation of the Pfaffian}
\label{sec:appendix_pfaffian_approach}
In this section we analytically derive the topological invariant of the two-wires system, namely a system composed by a single mode Sm coupled to a single mode Sc. We can write the Hamiltonian matrix as 
\begin{equation}
    \mathcal{H} = \begin{pmatrix} \mathcal{H}_\mathrm{Sm} & \mathcal{H}_\mathrm{t} \\
    \mathcal{H}^\dag_\mathrm{t} & \mathcal{H}_\mathrm{Sc}
    \end{pmatrix}   \,,
    \label{eq:bdg_complete}
\end{equation}
where the three terms read:
\begin{equation}
\label{eq:bdg_nw}
\mathcal{H}_\mathrm{Sm} = \qty(\frac{p_z^2}{2m_\mathrm{Sm}^*} - \mu_\mathrm{Sm}) \sigma_0\tau_z + \alpha_x p_z \sigma_y \tau_z \,,
\end{equation}

\begin{equation}
\label{eq:bdg_sc}
\begin{split}
\mathcal{H}_\mathrm{Sc} = \qty(\frac{p_z^2}{2m_\mathrm{Sc}^*} - \mu_\mathrm{Sc}) \sigma_0\tau_z + V_z \sigma_z \tau_0 + \Delta \sigma_0 \tau_x\,,
\end{split}
\end{equation}

\begin{equation}
\label{eq:bdg_t}
\mathcal{H}_\mathrm{t} = t_0 \sigma_0 \tau_z + t_z \sigma_z \tau_0\,.
\end{equation}

We proceed by rewriting the Hamiltonian in Majorana basis by using a unitary transformation $\mathcal{H}_\mathrm{MJ} =U \mathcal{H} U^\dag$ where the transformation matrix reads
\begin{equation}
    U = \frac{1}{\sqrt{2}} \begin{pmatrix} 
    +1 & 0 & 0 & -1 \\
    i & 0 & 0 & i \\
    0 & 1 & 1 & 0 \\
    0 & i & -i & 0
    \end{pmatrix} \otimes \mathbb{I}_2\,,
\end{equation}
where $\mathbb{I}_2$ is the $2\times2$ identity matrix.

Since the nanowire belongs to symmetry class $D$, the appearance of Majorana zero modes is characterized by a change of the sign of the topological invariant
\begin{equation}
    W_{\mathbb{Z}_2} = \mathrm{sgn} \qty[\frac{\mathrm{Pf} A(p_z = 0)}{\mathrm{Pf} A (p_z \to + \infty)}]
    \label{eq:topological_invariant}\,,
\end{equation} 
where $A = i \mathcal{H}_\mathrm{MJ}$ \cite{Tewari2012a}. The  matrix $A$ is  real and antisymmetric and thus the Pfaffian is well defined. Since the kinetic energy dominates for large $p_z$, the Hamiltonian approaches the one for quasi-free electrons, which have Pfaffian $\mathrm{Pf}A(p_z\to\infty) = 1$. Therefore evaluating Eq. \eqref{eq:topological_invariant} reduces to computing the Pfaffian for $p_z= 0$. Taking the limit $\Delta_0 \to 0$ we found the same condition described by Eq. \eqref{eq:self_energy_critical_lines}.

\section{Multiband superconductor coupled to a semiconductor}
\label{sec:appendix_multimode_sc}
In this section we derive expressions for the induced parameters in the Sm due to the coupling to a multimode Sc. We consider that, in the relevant energy scale, the band structure of the Sc can be described by $N_m$ modes distributed in an energy range $E_B$. For simplicity, we will take equidistant modes with energy difference $\delta \varepsilon = E_B/ N_m$. This is also consistent with the continuum limit since, for a 2D system, we can expect a density of transverse modes $g_m = (\delta\varepsilon)^{-1}$ constant in energy. We also define $\delta\mu_\mathrm{Sc}$ as the relative position of the Fermi energy in the Sc from the middle of the band.  We proceed by summing over the modes contributions following Eqs. (\ref{eq:induced_V1})--(\ref{eq:induced_Delta}) to derive the value of the induced terms in the effective Hamiltonian. In the limit $N_m = E_B/\delta\varepsilon \gg 1$ we get to 

\begin{equation}
    \tilde{V}_z^{(1)}(p_z) \simeq \frac{\pi}{2} g_m V_z (t_0^2 + t_z^2) 
    \frac{
        \cot\qty[\pi g_m \qty(i \Gamma + \delta\mu - \frac{p_z^2}{2 m^*_\mathrm{Sc}})]
      + \cot\qty[\pi g_m \qty(i \Gamma - \delta\mu + \frac{p_z^2}{2 m^*_\mathrm{Sc}})]
      }{i \Gamma}\,,
\end{equation}

\begin{equation}
    \tilde{V}_z^{(2)}(p_z) \simeq \pi g_m t_0 t_z 
    \qty{ 
      \cot\qty[\pi g_m \qty(i \Gamma + \delta\mu - \frac{p_z^2}{2 m^*_\mathrm{Sc}} )]
    - \cot\qty[\pi g_m \qty(i \Gamma - \delta\mu + \frac{p_z^2}{2 m^*_\mathrm{Sc}} )]
    }\,,
\end{equation}

\begin{equation}
    \tilde{\Delta}_0(p_z) \simeq \frac{\pi}{2} g_m \Delta_0 (t_0^2 - t_z^2) 
    \frac{ 
      \cot\qty[\pi g_m \qty(i \Gamma + \delta\mu - \frac{p_z^2}{2 m^*_\mathrm{Sc}})] 
    + \cot\qty[\pi g_m \qty(i \Gamma - \delta\mu + \frac{p_z^2}{2 m^*_\mathrm{Sc}})] 
    }{i \Gamma}\,.
\end{equation}

To simplify the expression, we can define the auxiliary functions 
\begin{equation}
\begin{split}
    F(x, y) &= \frac{\cot(i x + y) - \cot(i x - y)}{2} = \frac{\sin( 2 y)}{\cosh(2 x) - \cos(2 y)}\,, \\
    G(x, y) &= i \frac{\cot(i x + y) + \cot(i x - y)}{2} = \frac{\sinh(2 x)}{\cosh(2 y) - \cos(2 x)}\,. \\
\end{split}
\end{equation}

We finally get to compact expressions for the effective Hamiltonian terms:

\begin{equation}
    \tilde{V}_z^{(1)}(p_z) \simeq \pi g_m \frac{V_z}{\Gamma} (t_0^2 + t_z^2) G \qty[\pi g_m \Gamma, \pi g_m  \qty(\delta\mu - \frac{p_z^2}{2 m^*_\mathrm{Sc}})]
\end{equation}

\begin{equation}
    \tilde{V}_z^{(2)}(p_z) \simeq 2 \pi g_m t_0 t_z F \qty[\pi g_m \Gamma, \pi g_m  \qty(\delta\mu - \frac{p_z^2}{2 m^*_\mathrm{Sc}})]
\end{equation}

\begin{equation}
    \tilde{\Delta}_0(p_z) \simeq - \pi g_m \frac{\Delta_0}{\Gamma} (t_0^2 - t_z^2) G \qty[\pi g_m \Gamma, \pi g_m  \qty(\delta\mu - \frac{p_z^2}{2 m^*_\mathrm{Sc}})]
\end{equation}

This expression can be used to derive the Sm effective Hamiltonian and, with a simple linear search for the minimum eigenvalue, find the effective gap shown in Fig. \ref{fig:multi_modes_phasediagram}.

\end{widetext}
\end{document}